\documentstyle[preprint,tighten,aps,epsfig]{revtex}

\begin{document}
\title{ Parity-Violating $\Delta(1232)$ Electroweak Production:\\
Axial Structure and New Physics}
\author{S.J. Pollock
\thanks{Talk given at the ECT/CEBAF Workshop on N* Physics and
Nonperturbative QCD, Trento, May 1998. }
}
\address{University of Colorado, Boulder CO 80309}
\author{Nimai C. Mukhopadhyay}
\address{Rensselaer Polytechnic Institute, Troy, NY 12180}
\author{M. Ramsey-Musolf}
\address{University of Connecticut, Storrs, CT 06260}
\author{H.-W. Hammer}
\address{TRIUMF, 4004 Wesbrook Mall, Vancouver, B.C., Canada V6T 2A3}
\author{J. L\'\i u}

\sloppy
\maketitle

\begin{abstract}

We consider prospects for studying the parity-violating 
electroweak excitation of the $\Delta \left( 1232\right) $ resonance
with polarized electron scattering\cite{main}.  We discuss the
experimental feasibility and theoretical interpretability of such a
measurement as well as the prospective implications for hadron
structure theory. We also analyze the extent to which a PV
$N\to\Delta$ measurement could constrain various extensions of the
Standard Model.

\end{abstract}

If one measures a parity-{\em violating} (PV) asymmetry in
electroproduction of the $\Delta(1232)$ using longitudinally polarized
electrons on unpolarized nucleons, the resonant multipoles cancel in
the ratio, leaving behind a relatively clean and simple structure.  The
$N\to\Delta$ transition is appealing for a PV experiment, since the
$\Delta(1232)$ resonance is isolated from other nucleon resonances, the
production cross section is relatively large, and the transition is
pure isovector. This dominantly isovector nature eliminates
uncertainties associated with the unknown strangeness content of the
nucleon, and  yields a unique sensitivity to possible contributions
from additional heavy particles not appearing in the Standard Model
(SM).  $N\to\Delta$ offers the additional advantage that it only
couples strongly to one channel, $N\pi $.  This allows one to
treat unitarity issues quite rigorously\cite{1}.

In a recent paper\cite{main}, we examined this PV observable in some
detail.  The asymmetry has the following structure \cite{main,4,5}:
\begin{equation} 
\label{alr1} A_{LR}= {{N_+ - N_-}\over{N_++N_-}}=
\frac{-G_\mu}{\sqrt{2}}\frac{|q^2|}{4\pi \alpha }
	\left[ \Delta^\pi_{(1)} + \Delta^\pi_{(2)}+\Delta^\pi_{(3)}\right] ,
\end{equation} 
where $N_{+}$ ($N_{-}$) are the number of scattered electrons
for a beam of positive (negative) helicity electrons, $q^2$ tells
the square of the four-momentum transfer to the target; $\alpha $ and
$G_\mu$ are the fine structure constant and the Fermi constant.

The quantities $\Delta^\pi_{(i)}$ denote the three
primary contributions to the asymmetry discussed above.
$\Delta^\pi_{(1)}= 2(1-2{\sin^2\theta_{\scriptscriptstyle W}})$
includes the entire resonant hadronic vector-current contribution to
the asymmetry, at tree level in the SM.  This term contains no
dependence on hadronic form factors, owing to a cancellation between
terms in the helicity-dependent and helicity-independent cross
sections. The quantity $\Delta^\pi_{(2)}$ contains residual
contributions from non-resonant, hadronic vector-current isoscalar
backgrounds. The third term involves the
axial-vector $N\to\Delta$ coupling: $\Delta^\pi_{(3)}\approx
2(1-4{\sin^2\theta_{\scriptscriptstyle W}}) F(q^2,s)$ (plus 
axial vector background contributions.)  $F(q^2,
s)$ involves a ratio of electroweak response functions, $s$ is the
square of the total energy in the center of mass frame.

A precise measurement of $\Delta^\pi_{(1)}$ would provide a window on
physics beyond the SM.  We have found that a 1\% knowledge of
$\Delta^\pi_{(1)}$ would provide constraints roughly comparable to those
presently obtained from atomic PV.  A fairly demanding experimental
setup might be able to achieve this level, if the non-resonant
backgrounds and the axial contributions can both be understood at
roughly 25-30\% levels or better.  The third term, $\Delta^\pi_{(3)}$,
is extremely interesting from the standpoint of hadron structure. To a
good approximation, the function $F(q^2, s)$ is proportional to the
ratio of two transition form factors: $C^A_5/C^V_3$, where $V$ ($A$)
correspond to the hadronic vector (axial vector) current. This ratio is
the off-diagonal analog of the $G_A/G_V$ ratio in neutron
$\beta$-decay. A measurement of $\Delta^\pi_{(3)}$ could
correspondingly provide an opportunity to test low-energy consequences
of chiral symmetry, such as the off-diagonal Goldberger-Treiman
relation and its (small) chiral corrections \cite{HHM}.  In addition, a
determination of $C^A_5$ could provide tests of lattice and quark model
calculations and of the recipes proposed for correcting the vector form
factor discrepancies.

The most serious uncertainties appear in two guises: (i)
background contributions, contained in $\Delta^\pi_{(2)}$, and (ii)
hadronic contributions to electroweak radiative corrections including
two-boson exchange \lq\lq dispersion corrections" and corrections
induced by PV quark-quark interactions in the hadronic vertex. The
former enter the analysis of all three of the $\Delta^\pi_{(i)}$, while
the latter contribute to $\Delta^\pi_{(3)}$ only.  Although an estimate
of these hadronic PV corrections goes beyond the scope of the present
work, we emphasize the importance of performing such an estimate when
seeking to extract $C^A_5/C^V_3$ from $\Delta^\pi_{(3)}$.

For numerical estimates, we assume a plausible experimental
scenario\cite{Wel97b}, 1000 hours of 100\% polarized beam with a
luminosity of $2\times 10^{38}\text{ cm}^{-2}\text{ s}^{-1}$, solid
angle 20 msr, energy range for the outgoing electrons 0.2 GeV. The
figure of merit can easily be scaled for other assumptions. (For
example, our solid angle is grossly conservative for backward angle
experiments.) In Fig. 1a we show the $Q^2$-dependence of the axial term
for several electron energies.  We included a variety of
models\cite{main}, and the theory spread is commensurate with the
precision with which we anticipate one might realistically expect to
determine the axial term.  Consequently, a measurement of
$A_{LR}(N\to\Delta)$ may only be marginally useful as a discriminator
among models. Still, it would afford a determination of
$C_5^A/C_3^V$ at the level of the experimental-theoretical
discrepancies arising in the vector-current sector.  More detailed
numbers for various kinematics can be found in Table \ref{tableii}.
Fig. 1b shows the contributions of all three terms to the total
asymmetry at forward angles.  E.g., at $\theta=10^\circ$ and
$\epsilon=1$ GeV, we find $\Delta^\pi_{(2)}/\Delta^\pi_{(3)}\approx 6\%
$, and so even a large uncertainty in $\Delta^\pi_{(2)}$ has negligible
effect on an extraction of the axial term.  

In our calculations, we use a recent model-dependent estimate of the
background\cite{11}. At kinematics well suited for a determination of
$\sin^2\theta_{\scriptscriptstyle W}$, the vector-current background
contributes about 4-6\% of the total. Thus, a probe for new physics at
the 1\% level would require a theoretical uncertainty in the
background to be no more than 15-25\% of the total for
$\Delta^\pi_{(2)}$. Since the present model permits a 50\%
uncertainty in the background and still produces agreement with
inclusive EM pion production data, one could argue that a model
estimate of $\Delta^\pi_{(2)}$ is not sufficient for purposes of
undertaking a 1\% SM test. It appears that a model-independent
approach with an {\em experimental} isospin decomposition of the
EM pion production process offers the best hope for 
eliminating vector-current background uncertainties.

In summary, we have analyzed the PV $N\to\Delta$ transition asymmetry.
We considered the sensitivity of $A_{LR}$ to various scenarios for
physics beyond the SM -- such as leptoquarks, additional neutral gauge
bosons, and fermion compositeness -- as well as to transition form
factors of interest to hadron structure theory. After estimating the
precision with which $A_{LR}$ might be determined in a realistic
experiment, we estimated the scale of background effects. The use of
$A_{LR}(N\to\Delta)$ as a probe of hadron structure appears to be a
feasible prospect at present. A $\sim25\%$ determination of the
hadronic axial vector response could be carried out with realistic
running times. At reasonable kinematics for such a measurement, the
backgrounds appear to be sufficiently under control.  While a 25\%
determination of $C^A_5/C^V_3$ would not allow for a detailed
discrimination among model predictions, it would significantly improve
upon knowledge from charged current neutrino reactions and test model
predictions at the level of the theoretical-experimental discrepancies
arising in the vector current sector. A complete theoretical analysis
of $\Delta^\pi_{(3)}$, including effects of potentially large and
uncertain radiative corrections associated with hadronic PV, awaits a
future study.

\acknowledgments

The work was supported in part by the U.S. Department of Energy.
Thanks also to the ECT/CEBAF for their partial support at this
workshop.

\begin{table}
$$
\begin{array}{|ccc|ccccc|} \hline
\vphantom{\Biggl( }
E\left( {\rm GeV}\right) &
\theta_{\rm lab}(^o) &
Q^2 ({\rm GeV}^2) &
10^5 A_{\rm tot} &
\frac{\delta A_{\rm stat}}{A_{tot}}(\%) &
\frac{\Delta^\pi_{(2)}}{\Delta^\pi}(\%) &
\frac{ \Delta_{(3)}^\pi}{\Delta^\pi}(\%) &
\frac{ A_{(3)}}{\delta A_{stat}} \\
\hline
  .5 &  10. &  .002 &   -.03 &  45.9 & -.24  &20.4 &   .4 \\
 1.0 &  10. &  .020 &   -.22 &   5.9 & -.7   &11.0 &  1.9 \\
 4.0 &  10. &  .418 &  -4.17 &    .7 & -5.5  & 2.6 &  4.0 \\
\hline
  .5 &  90. &  .106 &  -1.24 &  10.9 & -2.81 &17.1 &  1.6 \\
 1.0 &  90. &  .641 &  -6.79 &   5.2 & -6.4  & 8.3 &  1.6 \\
 4.0 &  90. & 5.566 & -54.46 &  26.8 & -13.4  &  .8 &   .0 \\
 \hline
  .5 & 180. &  .157 &  -1.81 &  11.6 & -3.70  &15.5 &  1.3 \\
 1.0 & 180. &  .846 &  -8.85 &   7.5 & -7.0  & 7.1 &   .9 \\
 4.0 & 180. & 6.150 & -60.08 &  43.8 & -10.4  &  .6 &   .0 \\
\hline
\end{array}
$$
\caption{Our estimates of $A_{LR}$, experimental
statistical uncertainty for $A_{LR}$ (given our assumptions), 
vector-current backgrounds, contribution of axial multipoles,
and ratio of axial contribution to statistical uncertainty,
respectively, as functions of electron energy and scattering angle.  $Q^2$ is
calculated assuming we are on the $\Delta$ peak.
$\Delta^\pi=\Delta^\pi_{(1)}+\Delta^\pi_{(2)}+\Delta^\pi_{(3)}$, 
$A_{(3)}$ is the contribution to the asymmetry arising from
$\Delta^\pi_{(3)}$.}
\label{tableii}
\end{table}


\begin{figure}[htb]
\begin{center}
\psfig{file=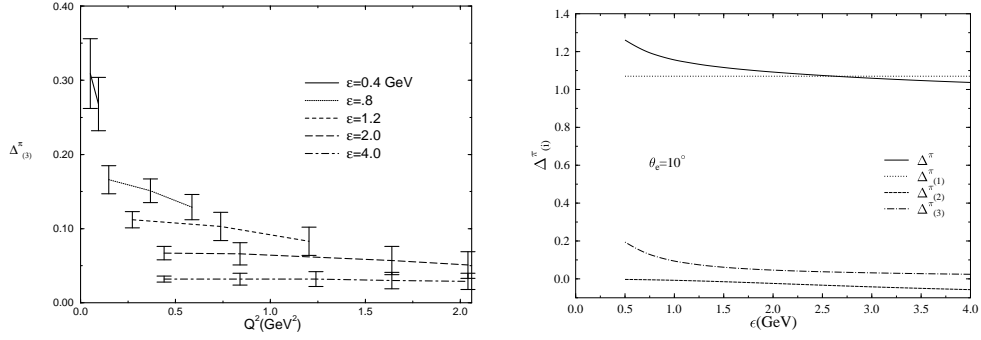,height=13cm,angle=-90}
\caption{a) 
$\Delta_{(3)}^\pi$ as a function of $Q^2$ for
different incident electron energies, $\epsilon$.
Error bars show a rough spread of theoretical results.
b) Contributions to the asymmetry from the $\Delta^\pi_{(i)}$
as a function of incident energy for $\theta
=10^\circ$. Here, $\Delta^\pi=
\Delta^\pi_{(1)}+\Delta^\pi_{(2)}+\Delta^\pi_{(3)}$.}
\end{center}
\end{figure}


\begin{figure}[htb]
\begin{center}
\psfig{file=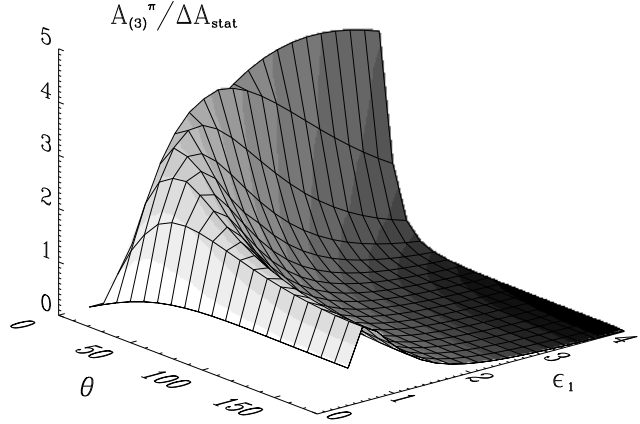,width=9cm}
\end{center}
\caption{ A 3-D plot of
$A_{(3)}/A_{\rm tot}\over \delta A_{\rm stat}/A_{\rm tot}$,
versus both incident energy and electron scattering
angle. (The shading is determined by the value of $\Delta_{(3)}^\pi$,
smaller values are shaded darker) }
\label{adoda-3d}
\end{figure}

\end{document}